\documentclass[letterpaper,twocolumn,10pt]{ilass}
\def\thepage{}

\usepackage{amssymb}
\usepackage{amsmath}
\usepackage{authblk}
\usepackage[T1]{fontenc}
\usepackage[utf8]{inputenc}
\usepackage[dvipsnames]{xcolor}
\usepackage{soul}
\usepackage{ifpdf}

\ifpdf
\usepackage[pdftex]{graphicx}
\else
\usepackage{graphicx}
\fi

\title{\vspace{-0.25in} {\small \em ILASS-Americas 31st Annual Conference on Liquid
                   Atomization and Spray Systems,
                   May 2021} \newline \newline
        \large {\bf Improved Methods for Mixing-Limited Spray Modeling} }
        
\author[1]{Majid Haghshenas\footnote{Corresponding Author: mhaghshenas@umass.edu}}
\author[1]{Peetak P. Mitra}
\author[2]{Chu Wang}
\author[3]{Fabien Tagliante}
\author[3]{Lyle Pickett}
\author[1]{David P. Schmidt} 
\affil[1]{Mechanical and Industrial Engineering Department,
                 University of Massachusetts Amherst, MA}
\affil[2]{Convergent Science Inc., Madison, WI}
\affil[3]{Sandia National Laboratories, Livermore, CA}

\date{ \normalsize  \centerline{\bf Abstract} \vspace{0.05in}
 \begin{minipage}{6.5in}
 \normalsize The realization that interfacial features play little role in diesel spray vaporization and advection has changed the modus operandi for spray modeling.  Lagrangian particle tracking has typically been focused on droplet behavior, with sub-models for breakup, collision, and interfacially-limited vaporization. In contrast, the mixing-oriented spray models are constructed so that gas entrainment is the limiting factor in the evolution of momentum and energy. In this work, a new spray model, ELMO (Eulerian Lagrangian Mixing-Oriented), is implemented in a three-dimensional CFD code with two-way coupling with the gas phase. The model is verified and validated with three canonical sprays including spray A, H, and G from the ECN database. 
 \end{minipage} \vspace{-0.25in}
}

\begin{document}

\ifpdf
\DeclareGraphicsExtensions{.pdf, .jpg}
\else
\DeclareGraphicsExtensions{.eps, .jpg}
\fi

\maketitle

\clearpage 
\pagenumbering{arabic}
\setcounter{page}{2}

\section*{Introduction}
Spray modeling in CFD has long been dominated by drop-oriented Lagrangian-Eulerian methods, where the stochastic nature of the algorithm provides efficient sampling of a sub-set of the actual drops \cite{schmidt2018} without the frustrations of numerical diffusion \cite{dukowicz1980}. The solution of the spray, which is statistically represented as  parcels, depends on the droplet breakup models, drag models, collision models, and turbulent dispersion models. The essence of the spray model was drop-oriented, while there could be partial consideration for the presence of neighboring drops in dense sprays \cite{balachandar2019}. This method was successful under many situations, and for this reason, Lagrangian-Eulerian modeling, such as Kelvin-Helmholtz Rayleigh Taylor (KH-RT), remains a mainstay of spray modeling with CFD. One major drawback of these models is that there are many adjustable constants that need to be tuned, each of which have an effect on model predictions. For example, $B_0$, $B_1$,  $C_t$ and $C_{RT}$  are the KH droplet size constant, KH breakup time constant, RT droplet size and RT breakup time constant respectively. 

Experimental studies conducted by Siebers et al. \cite{siebers1999scaling, naber1996effects, siebers1998liquid} contradict the basis of past spray models such as KH-RT. Siebers \cite{siebers1999scaling, naber1996effects, siebers1998liquid} concluded '\textit{the processes of atomization and the ensuing interphase transport of mass and energy at droplet surfaces are not limiting steps with respect to fuel vaporization in Direct Injection (DI) diesel sprays but rather they are limited by the turbulent mixing between phases}'. Moreover, a recent numerical study on spray evolution by Agarwal and Trujillo \cite{agarwal2018closer} questioned the validity of common spray models' assumed linking linear stability with primary atomization. 

Siebers \cite{siebers1999scaling} proposed a stand-alone steady jet one-dimensional simplified model for predicting liquid length for non-vaporizing sprays. Other studies, including Pastor et al. \cite{pastor20081d, pastor2011analysis}, Desantes et al. \cite{desantes2007evaporative, desantes20091d} and Musculus et al. \cite{musculus2009entrainment} extended the original formulation using the Eulerian control volume approach to add the physics of vaporization, model reacting flows, and include the end of injection transients with great success. These models assumed a radial profile of the spray contrary to the assumptions in the original work \cite{desantes2007evaporative, pastor20081d}. This is consistent with experimental and high-fidelity simulation observations for sprays. The typical profile for a spray during its evolution is shown in Figure \ref{fig:Ch5AssumedProfile, importance of alpha}. The spray starts off as a slug flow near the nozzle exit, and as it progresses downstream of the nozzle exit, acquires a radial profile due to aerodynamic interactions. Here $\alpha$ is the shape factor that allows the profiles to evolve from a uniform shape at the nozzle ($\alpha=\infty$) to a fully developed shape ($\alpha=1.5$)  far downstream \cite{musculus2009entrainment}. 

\begin{figure}[ht]
    \centering
    \includegraphics[width=0.85\linewidth]{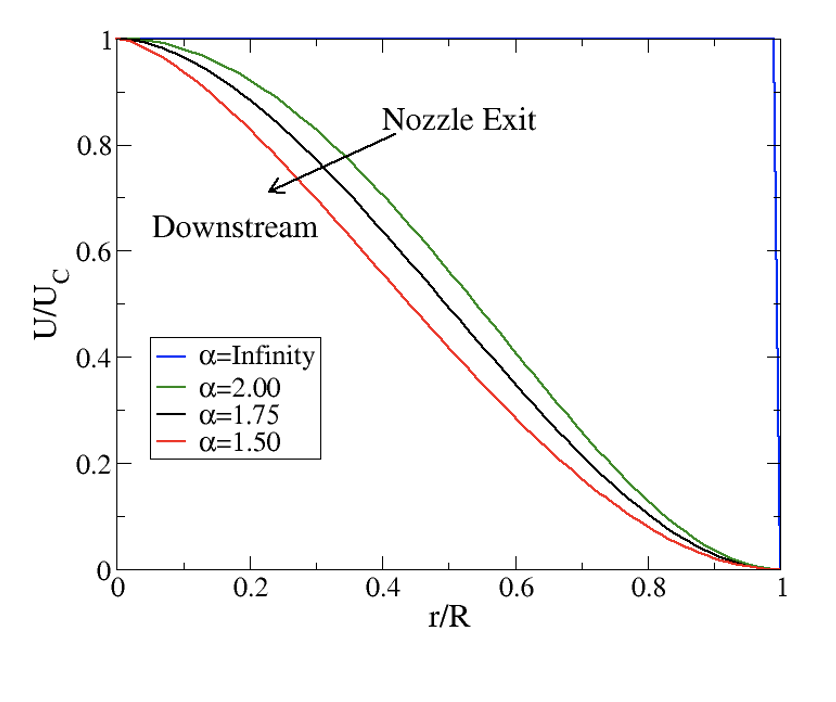}
    \caption{The spray profile behavior from nozzle exit to downstream. $U_C$ is centerline velocity and $R$ is radius of the spray profile.} 
    \label{fig:Ch5AssumedProfile, importance of alpha}
\end{figure}

\begin{figure}[ht]
    \centering
    \includegraphics[width=0.85\linewidth]{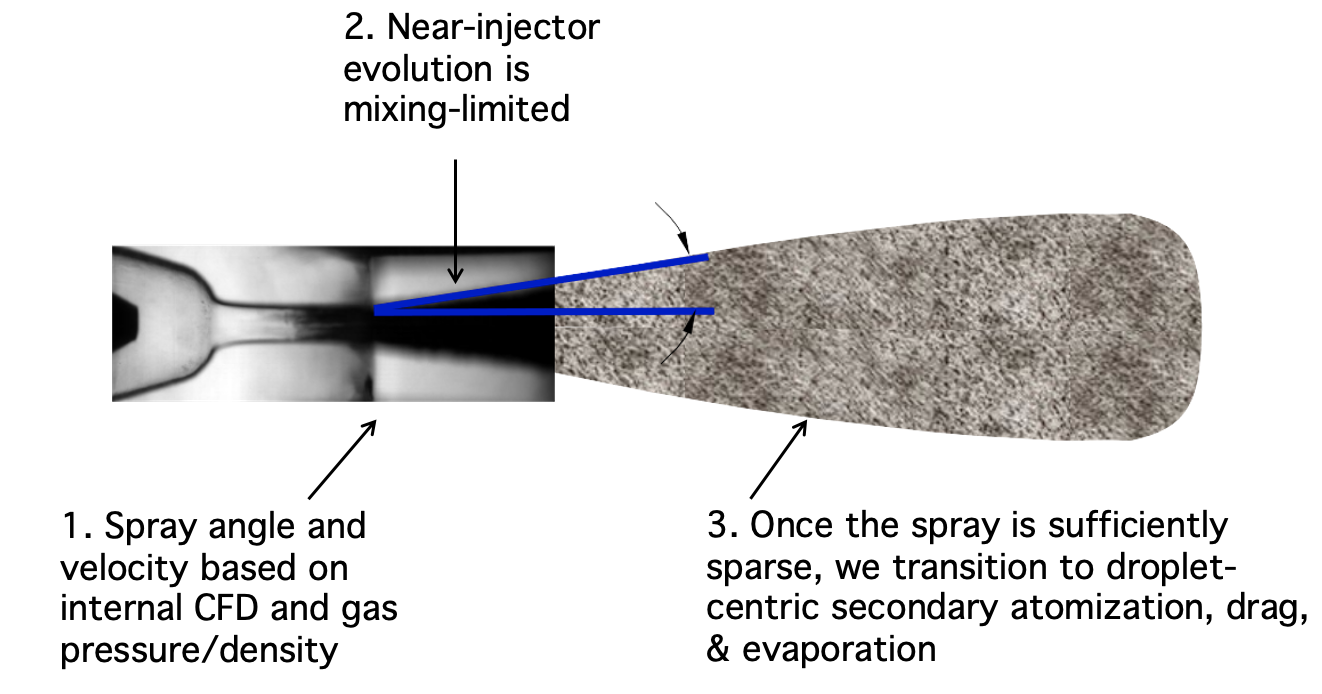}
    \caption{The ELMO model idealization is shown in this figure. In the near nozzle region, the ELMO operates under the mixing limited assumption. When the spray becomes sufficiently dilute, the ELMO transitions to standard LE and its droplet centric approach.}
    \label{fig:Ch5ELMOOverallSchematic}
\end{figure}

Most of the mixing-oriented models were formulated to be used only as stand-alone models and not be coupled with a multidimensional CFD solver. There have been several implementations of reduced-order Mixing-limited models in CFD codes in recent years \cite{abani2008,yue2017,yue2019,perini2021investigation}, however, these promising works assumed a constant spray angle and constrained the mass and momentum exchange to be mostly predetermined. The Eulerian Lagrangian Mixing Oriented (ELMO) model presented here is a fully coupled model that is well-suited to time-varying spray angle. The inputs of the ELMO model are informed by the internal flow and experimental observations, including discharge rate and spray angle. The near-nozzle spray evolution is based on the more accurate mixing-limited hypothesis inspired by previous efforts in this area \cite{siebers1999scaling, desantes2007evaporative, pastor20081d, musculus2009entrainment}. ELMO is coupled to a CFD solver, and in the dilute region downstream of the nozzle this model transitions to a standard Lagrangian Eulerian treatment with its droplet centric formulation. A schematic is shown in Figure \ref{fig:Ch5ELMOOverallSchematic}.

The advantages of this approach include: 1) The model can benefit from sophisticated transient internal models, informing the inputs for the external spray modeling paradigm. 2) The near nozzle behavior is consistent with the modern mixing limited hypothesis, therefore expected to be more accurate. 3) In the downstream region, the model employs conventional droplet models (collisions, drag) where they work best \cite{pai2009comprehensive}. 4) Since the spray angle is an input to the model, it can be more effectively modeled than in the current state-of-the-art. Siebers \cite{siebers1999scaling} shows the spray angle to be transient during high/low needle lift. Most of the current modeling approaches ignore this transient, yet important, behavior.

The Musculus-Kattke model \cite{musculus2009entrainment} is used as the starting point for the numerical formulation for the ELMO model. While the previous implementation is in an Eulerian frame of reference, the ELMO model is in the corresponding Lagrangian frame of reference. This essentially provides the capability to use the Lagrangian 'capsules' as self-contained liquid fuel parcels that entrain ambient air during the translation process along the spray axis. The capsule involves a moving Lagrangian parcel over an underlying Eulerian mesh. This capsule therefore has two phases, the liquid fuel and gas, including ambient air as well as vaporized fuel due to vaporization caused by the entrainment of hot ambient air. The in-capsule vaporization is modeled using the Desantes \cite{desantes2007evaporative} formulation based on equilibrium evaporative physics. Therefore, the ELMO model is built on assumptions of inertial and thermal equilibrium, limited only by the entrainment rate of surrounding ambient air.

In the following sections, first, the numerical methodology of ELMO method is presented. Next, the model is verified against Musculus-Kattke and Desantes models for kinematics and thermodynamics respectively. And finally, validation results are presented based on standard test cases from Engine Combustion Network \cite{ECNweb}, including spray A, H, and G. 

\section*{Methodology}
In this section we will introduce the detailed discussion about the numerical methods for the formulation of the ELMO model, including some of the assumptions.

\subsection*{The capsule formulation}

The Eulerian formulation in Musculus Kattke \cite{musculus2009entrainment}, while suitable for a standalone spray prediction model, is converted into its corresponding Lagrangian formulation. This is done to use the available capabilities of the underlying CFD code CONVERGE \cite{senecal2014grid, CONVERGE} in handling Lagrangian parcels. The ELMO model has a few assumptions:

\begin{itemize}
    \item The mass of the fuel in a capsule remains constant throughout the evolution of the transient, up to the point of transition. It can be represented mathematically as:
        \begin{equation}
            \frac{dm_{f}}{dt} = 0
            \label{eq:const_mass_f}
        \end{equation}
        
        where $m_f$ is the fuel mass
        
    \item The momentum of the capsule remains constant. This translates to the capsule velocity slowing down in proportion to the mass of the ambient air entrained. Mathematically, at an individual capsule level this can be represented as:
        \begin{equation}
             \frac{dM}{dt} = 0
        \end{equation}
        
        where $M$ is the momentum
        
    \item Each capsule starts off with its spray angle at birth. This spray angle dictates the shape and entrainment within the capsule and remains constant throughout the capsule's life, even as the spray angle of injection changes over time. 

    \item Underlying all of the equations (introduced later in this section) are assumptions of equilibrium. Specifically, the phases all move at the same velocity and experience equal pressure. All phases are in thermal equilibrium as well. The capsule is in thermodynamic equilibrium, with an assumption of locally homogeneous flow. Therefore, the liquid and gas are at the same temperature.
\end{itemize}

Applying the Reynolds Transport Theorem (RTT), we convert the Eulerian MK control volumes \cite{musculus2009entrainment} into the Lagrangian ELMO capsules.

\begin{equation}
    \frac{dm_{f}}{dt} = \frac{\partial}{\partial t} \int_{V} \rho_{f} \bar{X_{f}}dV + \int_{CS} \rho_{f}\bar{X_{f}}(\vec{u} - \vec{u_{r}})dA
\end{equation}

By design, the left hand side of the equation is set to zero. The volumetric integral time is also zero due to fuel mass conservation principles, which reduces the above equation to

\begin{equation}
    \int_{FS}\bar{X_f}\vec{u}.dA - \int_{FS}\bar{X_f}\vec{u_r} \cdot dA = 0
\end{equation}

which further can be reduced to:

\begin{equation}
    \vec{u_r} = \beta \vec{u}
\end{equation}

where $m_f$ is the mass of fuel, $u$ is the axial velocity, $u_r$ is the control surface velocity, $\beta$ is the shape factor of fuel volume fraction, $\rho_f$ is the density of fuel, $\bar{X_f}$ is the fuel volume fraction, $CS$ is the control surface and $FS$ is the front surface. The term $\beta$ accounts for the shape of the fuel volume fraction and velocity profiles, ranging from $\beta = 1$ for a uniform profile to approximately $\beta = 2$ for fully developed jets and is based on the representation shown in Figure \ref{fig:Ch5AssumedProfile, importance of alpha} for $\alpha$. For details about the shape factor term, please refer the original source of this work \cite{musculus2009entrainment}.

Similarly, applying the RTT, the momentum formulation can be represented as:
\begin{equation}
    \frac{dM}{dt} = \frac{\partial}{\partial t} \int_{V} \bar{\bar{\rho}} \vec{u}dV + \int_{CS} \bar{\bar{\rho}} \bar{\bar{u}}(\vec{u} - \vec{u_{r}}.n)dA
\end{equation}

per the design of the capsule concept and in the absence of a pressure gradient, the equation above reduces (approximately) to 

\begin{equation}
    \frac{\partial }{\partial t}\bar{\bar{u}}m = 0
\end{equation}

where $M$ is the momentum of the capsule, $\bar{\bar{\rho}}$ is the mixture density of the capsule, $\vec{u}$ is the axial velocity, $\bar{\bar{u}}$ is the cross-sectionally averaged axial velocity, $u_r$ is the control surface velocity and $m$ is the total mass of capsule. 

\begin{figure}[h]
    \centering
    \includegraphics[width=2.60in]{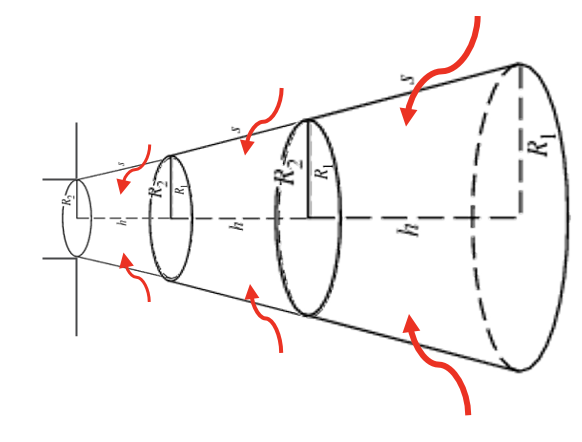}
    \caption{The schematic of the ambient air entrainment in the capsule.}
    \label{fig:Ch5CapsuleEntrainment}
\end{figure}

Since the fuel mass in a capsule remains constant, it can either exist in a liquid or vapor (gas) phase. The growth in volume (and mass) of the capsule at successive time instants is accounted by the entrainment of the ambient air (schematic shown in Figure \ref{fig:Ch5CapsuleEntrainment}). Therefore the mass and volume conservation equations have the following functional form:

\begin{equation}
    m = m_l + m_v + m_a
\end{equation}

\begin{equation}
    V = V_l + V_v + V_a
\end{equation}

where the subscripts $l$, $v$, and $a$ are liquid fuel, vapor, and non-condensible gas respectively. The presence of the non-condensible gas in the capsule necessitates calculating phase interactions with the underlying Eulerian CFD control volume. This approach will be detailed in the coming sections.

The increase in capsule mass due to the entrainment of air can be represented as

\begin{equation}
    m^{t+1}_{capsule} = m^t_{capsule} + \Delta m_{air}
\end{equation}

In the present implementation, the effect due to entrained momentum is ignored, similar to the original work \cite{musculus2009entrainment}. This therefore leads to a constant momentum for each capsule, with decelerating capsule velocity. Like the implementation of Musculus and Kattke, the consideration of a radial velocity and density profile is included here, which has implications for momentum. This radial profile is manifested by the term $\beta$ such that the velocity is higher along the spray axis, where most of the mass is concentrated. As the capsule advances in time, the position is updated by a simple explicit Euler time integration given by:

\begin{equation}\label{cap3}
    z^{t+1} = z^t + \beta \bar{u}\Delta t
\end{equation}

where $z$ is the axial location of the capsule, and $\bar{u}$ is the capsule velocity at the location $z$. While the capsule volume increases radially, its axial extent diminishes as the control surfaces of the capsules are advected at different velocities based on their location along the spray axis.  This same reduction in velocity represents a deceleration of the center of mass of each capsule. The derivation of the front side (fs) and back side (bs) location of each capsule begins with the assumption that these capsule boundaries lie halfway between the centroid $z$ of a capsule at its current position and the next position for the front side or halfway between the current centroid location and previous location for the back side. Expressing this mathematically:

\begin{equation}\label{cap1}
    fs^t = \frac{z^t + z^{t+1}}{2}
\end{equation}

\begin{equation}\label{cap2}
    bs^t = \frac{z^t + z^{t-1}}{2}
\end{equation}

In order to make the train of capsules continuous, we constrain the back surface of the leading capsule to match with the front surface of the trailing capsule.

Once the locations of the front and back sides are updated, the geometry of the capsule is fully determined and the kinematics of the advection process is complete. Once the axial extents of the capsules are known, the other features such as velocity, volume can be easily calculated.

\subsection*{Thermodynamic formulation}

To model the process of vaporization, we follow the analysis first proposed by Desantes et al.\cite{desantes2007evaporative}. Desantes et al.\ extended  a one-dimensional non-evaporative spray model to take into account the effects of evaporation under direct-injection diesel engine conditions. Their approach is based on the assumption that evaporation is limited by the mixing process between fuel and ambient gas, and makes use of non-ideal gas state relationships for the description of the whole spray. Their model has been shown to accurately predict the influence of fuel type and both ambient and injection conditions on liquid spray penetration. The hot entrained ambient air is the primary driver of the vaporization process. The local thermodynamic equilibrium in the capsule depends on the mass fractions. Mathematically, they are represented as below.

\begin{equation}
    Y_f = Y_{f,l} + Y_{f,v}
\end{equation}

\begin{equation}
    Y_a = 1 - Y_f
\end{equation}

where suffixes $f$, $a$, $l$, $v$ correspond to fuel, air, liquid, and vapor phase respectively. While in the original work, non-ideal state relationships were considered, in this implementation we assume all gases to be ideal. This greatly simplifies our calculations without appreciable differences between predictions. The local mixture density is calculated using an ideal mixture assumption

\begin{equation}
    \rho = \frac{1}{\frac{Y_{f,l}}{\rho_{f,l}} +\frac{Y_{f,v}}{\rho_{f,v}} + \frac{Y_{a}}{\rho_{a}} }
\end{equation}

where $Y_{i}$ is the mass fraction of the mixture $i$ component and $\rho_{i}$ is the density for the pure component $i$ at the mixture temperature $T$ and total pressure $P$. The local mixture enthalpy $h$ is also calculated through an ideal mixture assumption as

\begin{equation}
    h(T) = Y_{f,l} \cdot h_{f,l}(T) + Y_{f,v} \cdot h_{f,v}(T) + Y_{a} \cdot h_{a}(T)
\end{equation}

where $Y_{i}$ is the mass fraction of the mixture $i$ component and $h_{i}$ is the enthalpy for the pure component $i$ at the mixture temperature $T$.

Due to the liquid-vapor equilibrium, the composition and temperatures of phases are coupled  i.e., for a given $Y_f$, the proportion of liquid or evaporated fuel is not known, nor is the mixture temperature. It can however be solved iteratively, using the formulation below

\begin{equation} \label{eq1}
\begin{split}
 & \frac{Y_{f,v}}{1-Y_f}  =  \\
 & = \frac{h_a(T_{a,\inf} - h_a(T) - \frac{Y_f}{1-Y_f}[h_{f,l}(T)-h_{f,l}(T_{f,0})]}{\Delta h_v(T)}
\end{split}
\end{equation}
In this equation, if $Y_f$ is known the right hand side terms depend only on local temperature (T), and the left hand side on $Y_{f,v}$. This equation fully determines the thermodynamic equilibrium state of the capsule. The equilibrium equation can be written in terms of mass fractions and molecular weights as

\begin{equation}\label{eq2}
    \frac{Y_{f,v}}{1-Y_f} = \frac{MW_f}{MW_a}\frac{1}{\frac{p_a}{p_v}-1}
\end{equation}

where $MW_f$ is the molecular weight of the fuel, and $MW_a$ is the molecular weight of the ambient air. This helps to estimate the $Y_{f,v}$ term and therefore solve the equation \ref{eq1} using a root finding method to determine the temperature.

The ideal gas assumption necessitates using corrections (Poynting) in calculating vapor pressures \cite{desantes2007evaporative}.

\begin{equation}
    p_v = p^{0}_v*exp[V_l \frac{p_a-p^{0}_v}{R_{gas}T}]
\end{equation}

where the uncorrected vapor pressure is $p^{0}_v$, $R_{gas}$ is the gas constant, $p_{a}$ is the ambient pressure. This corrected vapor pressure is used in equation \ref{eq2}. 

A two-way coupling mechanism based on Monte Carlo (MC) integration is used to sample thermodynamic variables within the capsule geometry. This MC mechanism is also used to provide gas phase source term, as a result of vaporized fuel, to the Eulerian CFD mesh. A key feature of this implementation is the presence of gas in both the capsule (to account for effects of vaporization and mass/momentum) and the Eulerian mesh. To simplify our calculations, we define the gas phase gain equal to the loss of the liquid fuel. This evaporated fuel, now in the gas phase, is used in the source term for the mass. Further, the energy loss of the capsules at each time step is deposited as the energy source into the gas solution of the underlying gas cell. 

\subsection*{Coupling formulation}

\begin{figure}[h]
    \centering
    \includegraphics[width=2.2in]{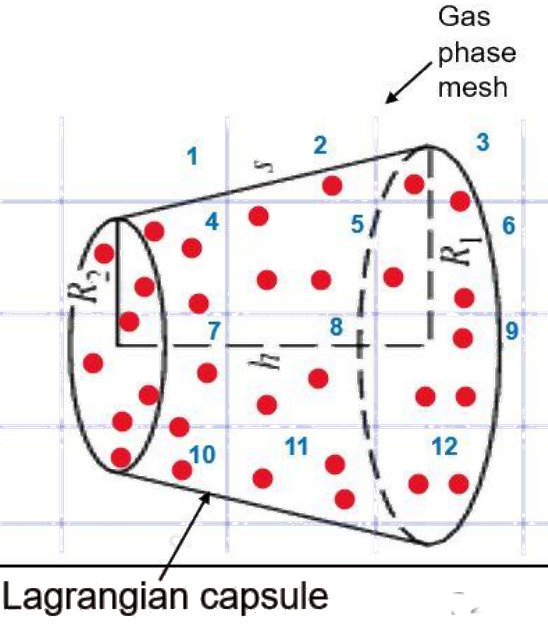}
    \caption{The schematic of a limited number of points distributed randomly within the capsule geometry. Blue numbers represent cell indices. The underlying Eulerian mesh computes the overlapping number of points within each cell and distributes the source terms.}
    \label{fig:Ch5CapsuleMesh}
\end{figure}

The source terms must be distributed to the underlying Eulerian mesh. Depending on the capsule spatial and radial extent, and the underlying mesh resolution each capsule may overlap with numerous gas phase cells. A Monte Carlo (MC) integration method is used to distribute the source terms to each gas phase cells. This coupling using MC is a two-way coupling, as the gas phase properties are obtained from the capsule advection equations. A single capsule may overlap with multiple mesh cells. This overlap complicates the redistribution of source terms to the underlying mesh. A simple strategy is based on the same point finding algorithm used to determine in which cell a Lagrangian parcel might reside is used. In this algorithm for a given capsule, for each of these Monte Carlo points, a fraction of $1/N$ of the source term is deposited into the cell which contains the point. For example, in the Figure \ref{fig:Ch5CapsuleMesh}, cell \# 8 has two randomly generated MC points coinciding with it. Therefore, $2/N$ of the total mass/momentum/energy in the capsule is added to the gas phase source term. This stochastic approach obviates the need for calculating expensive geometric intersections. These interior points are used only for distributing source terms. 

\subsection*{Transition}

As the capsule traverses along the axial direction, it ingests more ambient gas in the form of entrainment. This causes vaporization as the ambient gas temperatures are on average higher than the fuel temperature. This combined effect reduces the liquid fuel in the capsule, making the capsules diluted at a certain location downstream. At this point the Lagrangian capsule approach is not further needed, and the capsule is ready to be transitioned to Lagrangian parcels consisting of very small droplets. This transition has several advantages including making use of existing frameworks to capture parcel impacts on cylinder walls. In the current formulation, the capsules only traverse in the direction of the initial spray, but post transition parcels are not limited by this over-simplification. The flexibility of the post-transition parcels will help in the model's ability to successfully capture the phenomena of spray bending, as is observed in the case of multi-injector nozzles.

The criterion for transition in this current implementation is based on three factors:

\begin{itemize}
    \item The liquid volume fraction is below 0.005
    \item The vapor volume fraction of the total fuel volume fraction is above 99\%
    \item The capsule's average velocity is lower than 10 $m/s$
\end{itemize}

During transition, the remaining liquid mass in the capsule is proposed to be divided equally into a predetermined number of parcels at locations generated randomly within the capsule's geometrical extent. Since the capsule geometry is small, of the order of a few Eulerian mesh cells, the random distribution of parcels is not expected to make a great difference. The droplet size is based on a critical Weber number of 6, as indicated by experiments \cite{pitsch2006large}. Once the transition to Lagrangian parcels is complete, the parcels evolve with typical spray models appropriate for secondary atomization of dilute sprays.

\subsection*{Overall approach}

The kinematic equations are solved first using the same time-step ($\delta t$) as the CFD solver. The discharge rate curve is interpolated based on the overall timestep of the solution. The discharge rate determines the mass of fuel in each capsule upon initialization. The capsules are initialized at the nozzle exit, with an initial spray angle, and an axial length equal to the velocity (at birth) times $\delta t$. The capsules then evolve in the domain using equations \ref{cap3}. The new positions determine the volume of the capsule as the one-dimensional implementation is projected onto three dimensions using simplified assumptions of radial profiles. The entrained ambient air mass is calculated and the new velocity determined by conserving the capsule momentum. The solution of the thermodynamic state using the relationships discussed, equations \ref{eq1} and \ref{eq2}, requires iterative solution due to the coupling of the variables. Once the new thermodynamic state is found, the capsule calculations are concluded. At the end of each cycle of calculations, capsules are examined against the transition criterion and the ones meeting it are transitioned into smaller Lagrangian parcels.

\section*{Results}
\begin{figure}[ht]
 \centering
 \includegraphics[width=0.9\linewidth]{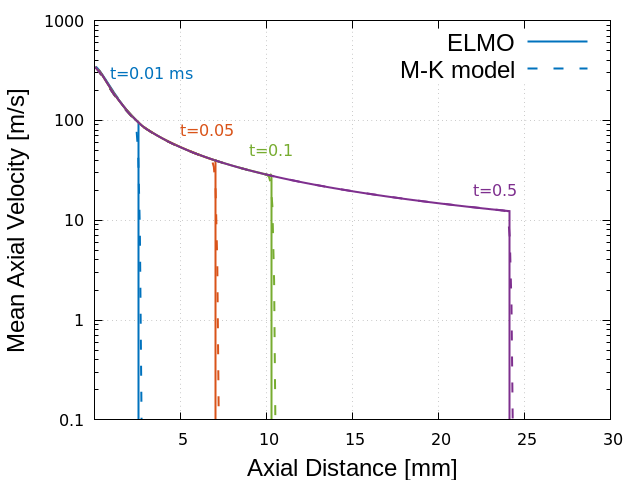}
 \caption{ Fuel jet penetration compared to Musculus-Kattke model \cite{musculus2009}.}
 \label{fig:MK_Ver}
\end{figure}

The results presented here comprise two categories:  verification and validation. Verification provides evidence that the kinematics and thermodynamics equations are being solved correctly, and validation provides evidence that the model represents actual sprays. Verification of the results begins with a spray penetration comparison to the Musculus-Kattke model \cite{musculus2009entrainment} in order to show that the kinematics implementation of the ELMO model is being solved correctly. The test case is based on the Engine Combustion Network (ECN) Spray A \cite{ECNweb} without vaporization, and the result is shown in Fig.\ \ref{fig:MK_Ver}. Mean axial velocity of fuel is plotted versus axial distance at different times of the solution to check the spray evolution. The prediction of ELMO model matches with the result of M-K model. 

\begin{figure}[ht]
 \centering
 \includegraphics[width=0.9\linewidth]{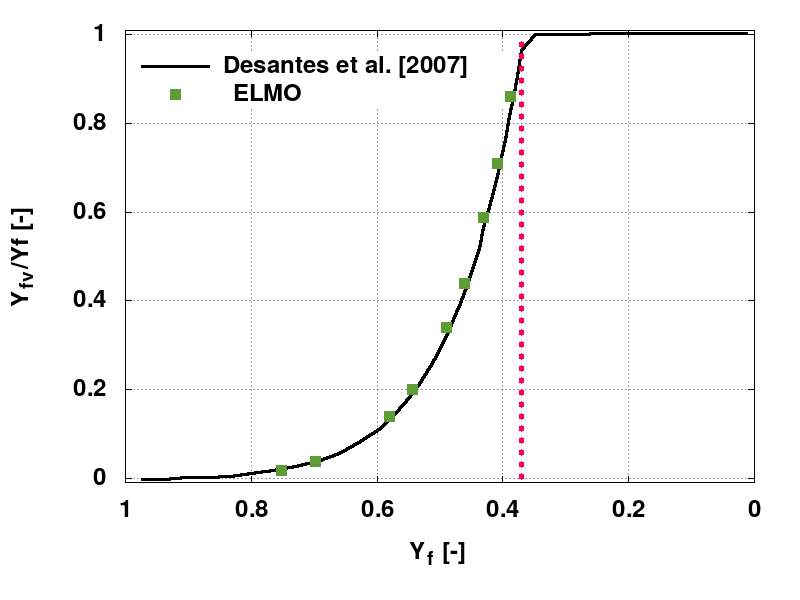}
 \caption{With a static, undisturbed flowfield, the current implementation reproduces the results of Desantes et al.  The abscissa represents the mass fraction of fuel and the ordinate is the fraction of fuel in the vapor phase.}
 \label{fig:DesantesComparison}
\end{figure}

Next, the thermodynamic implementation was tested as a verification measure to ensure that it was faithful to the thermodynamics of Desantes et al. \cite{desantes2007}. The test case is Spray A, while the source terms are switched off to make the ELMO code see a uniform ambient condition undisturbed by the spray. The results are shown in Fig.\ \ref{fig:DesantesComparison}. Mass fraction ratio of vapor to liquid fuel is plotted versus fuel mass fraction. It shows the vaporization progress versus the gas entertainment into the fuel capsule. The fuel is completely vaporized at a fuel mass fraction slightly less than 0.4.

For validation, three standard spray cases from Engine Combustion Network (ECN) were considered: Spray A, Spray H, and Spray G \cite{kastengren2012} \cite{duke2017} \cite{kosters2016}.  The validation includes global spray quantities, such as vapor penetration, radial fuel distribution, and gas motion. The ELMO model was used without any adjustment to the model parameters in all validation cases. For all the simulations, the based grid size is 2mm, while embedded grid refinement around the nozzle and adaptive mesh refinement (AMR) assist the simulation. Unresolved sub-grid scale features were modeled with RNG $k-\epsilon$ turbulence model. Taylor Analogy Breakup (TAB) \cite{orourke1987} and No Time Counter (NTC) \cite{schmidt2000} are considered as the breakup and collision models for the transition parcels. Also, the Fr{\"o}ssling \cite{frossling1938} method is employed as the vaporization model.  

\begin{figure}[ht]
    \centering
 a){{\includegraphics[width=3 in] {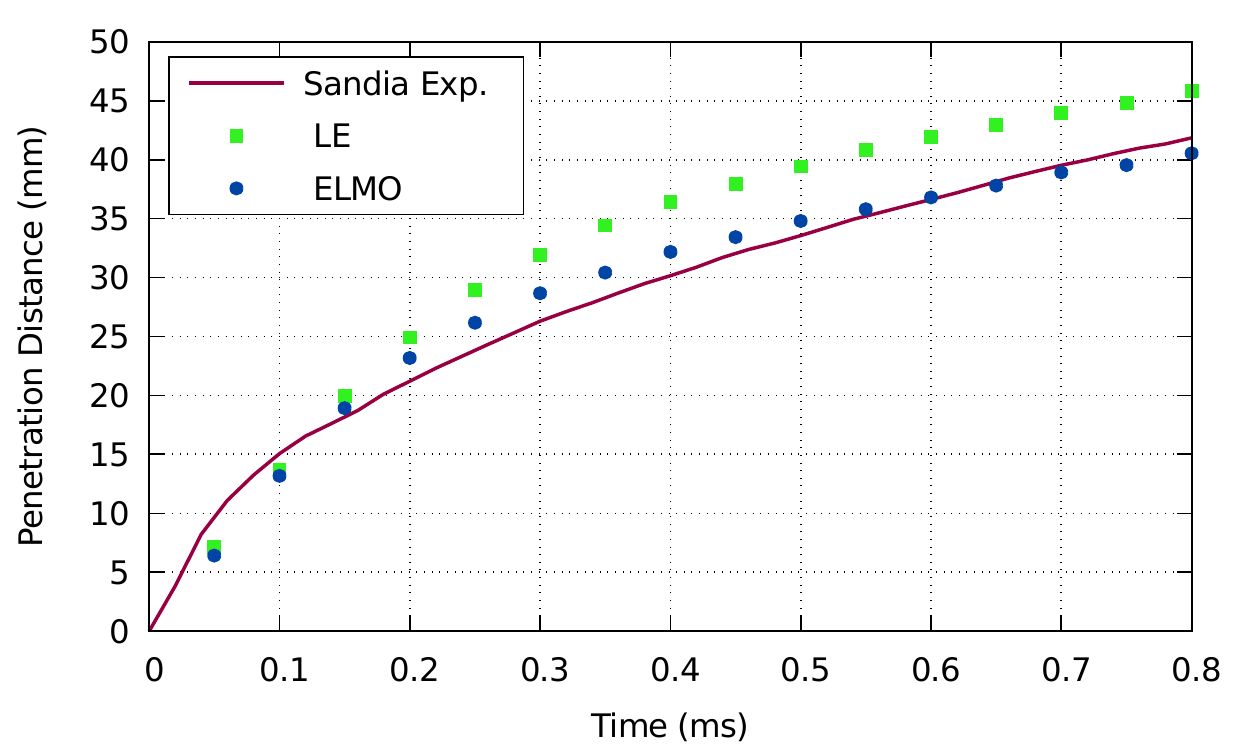} }}

 b){{\includegraphics[width=3 in]{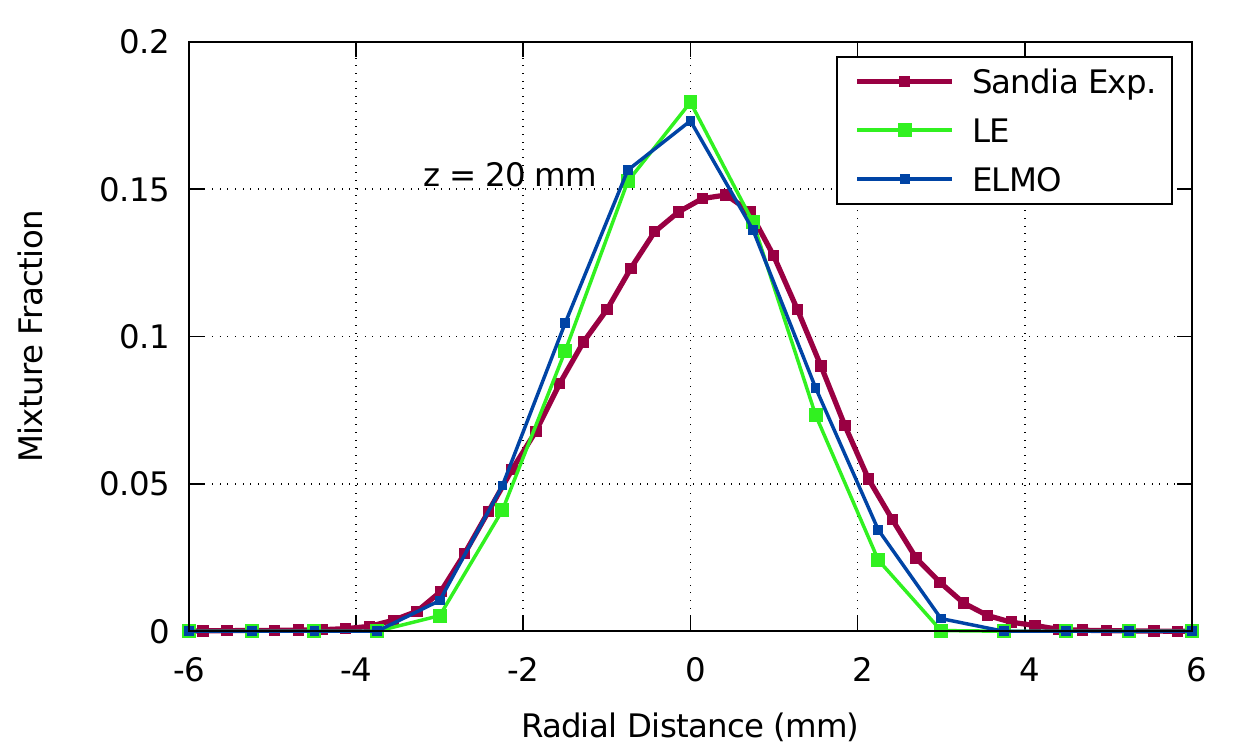} }}%
 c){{\includegraphics[width=3 in]{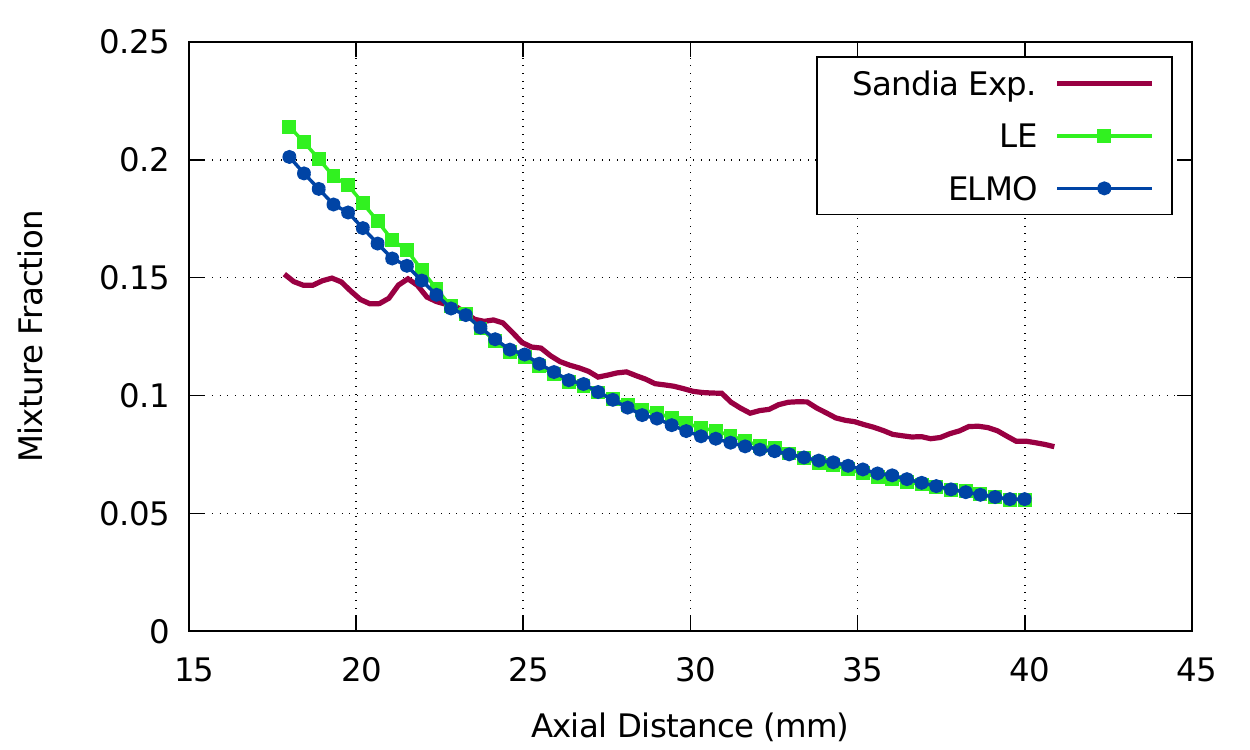} }}%
    \caption{Spray A simulation compared against experimental result of Pickett et al \cite{pickett2011}. a) vapor penetration, b) transverse and c) longitudinal mixture fraction distribution.}%
    \label{fig:SpA}%
\end{figure}

The first validation case is Spray A and the injection parameters can be found in the reference \cite{ECNweb}. For this case, spray cone angle is 21.5\textdegree and the injection coefficient of area is 0.98. The initial turbulent kinetic energy and dissipation are 5.02E-4 $m^2/s^2$ and 0.0187 $m^2/s^3$ respectively. The predictions of the ELMO model were compared to experimental data from Pickett et al \cite{pickett2011}.  Figure \ref{fig:SpA} shows the penetration versus time (a), the transverse distribution of fuel mass at 20 $mm$ downstream (b), and longitudinal distribution of fuel mass from 18 $mm$ to 40 $mm$ downstream of injection at the end of injection. For comparison, Lagrangian-Eulerian (LE) simulation results are also included in Fig.\ \ref{fig:SpA}.  The model inputs were specified as a part of an example case that comes with CONVERGE 2.4. The penetration is predicted by the ELMO model without any model adjustment.  Vapor penetration result by ELMO is in good agreement with the experiment and relatively outperformed the LE solution. The transverse and longitudinal profiles shown in Fig.\ \ref{fig:SpA} (b) and (c) permit comparison of the fuel dispersion. The fuel is entirely vaporized well before these measurement locations.  ELMO, similar to LE, predicts the bell-shaped distribution with the approximate width and height of the experimental measurements. For the longitudinal profile, ELMO shows some over-prediction compared with the the experimental data at locations near the nozzle, while it switches to under-prediction as we move downstream. There is a similar trend in the LE solution. It worth noting that fuel capsules evaporate entirely within a few millimeters of injector exit, and other factors such as turbulence and evaporation models could affect the downstream results. 

\begin{figure}[ht]
    \centering
 a){{\includegraphics[width=3 in]{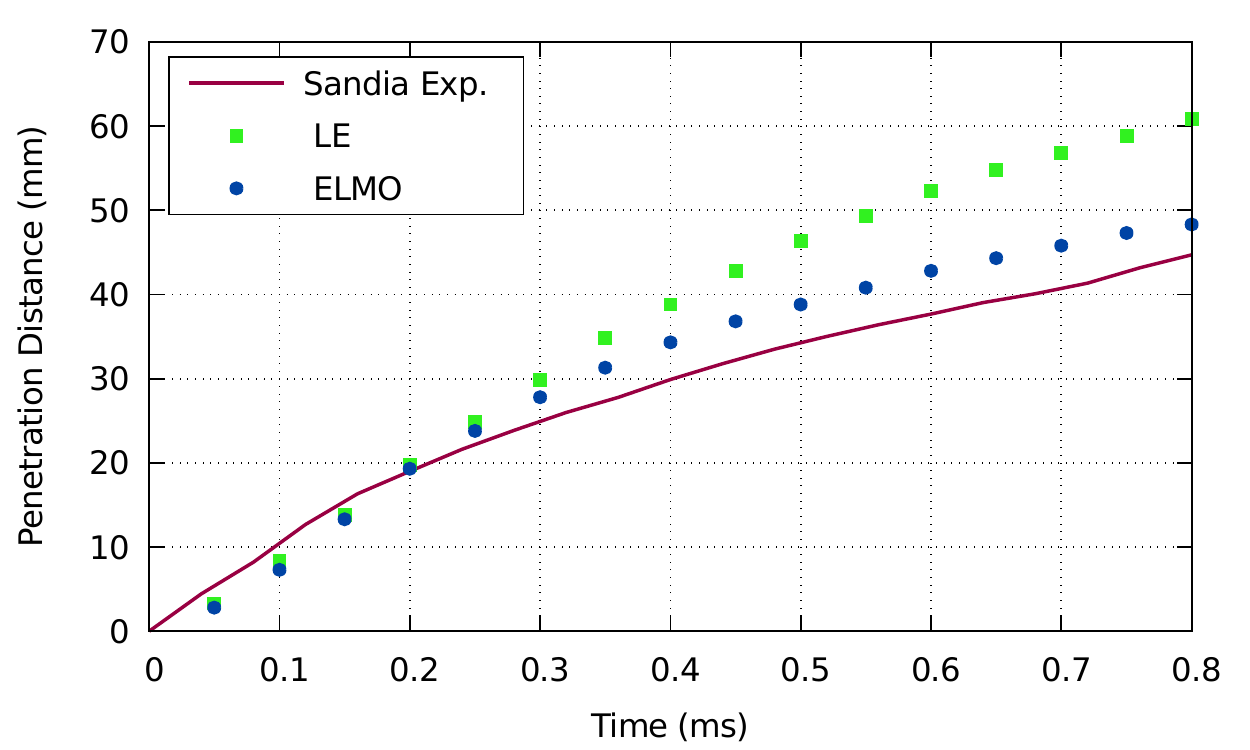} }}%
 b){{\includegraphics[width=3 in]{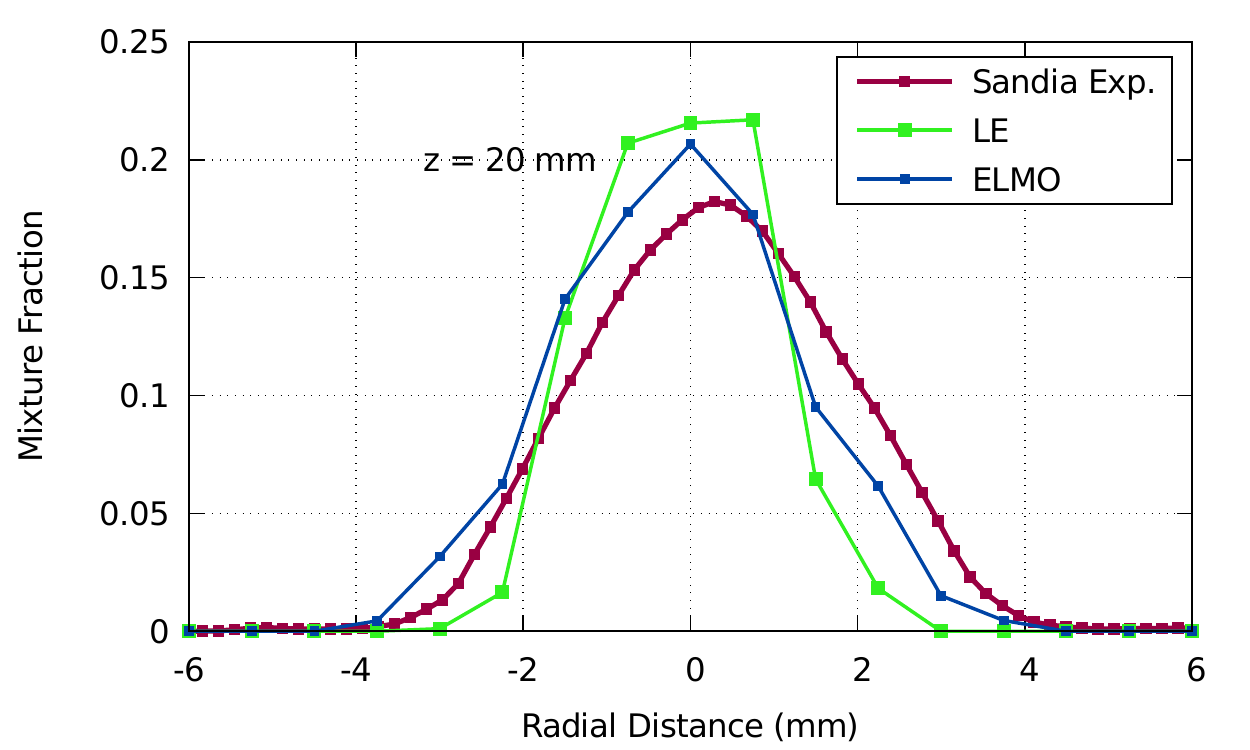} }}%
 c){{\includegraphics[width=3 in]{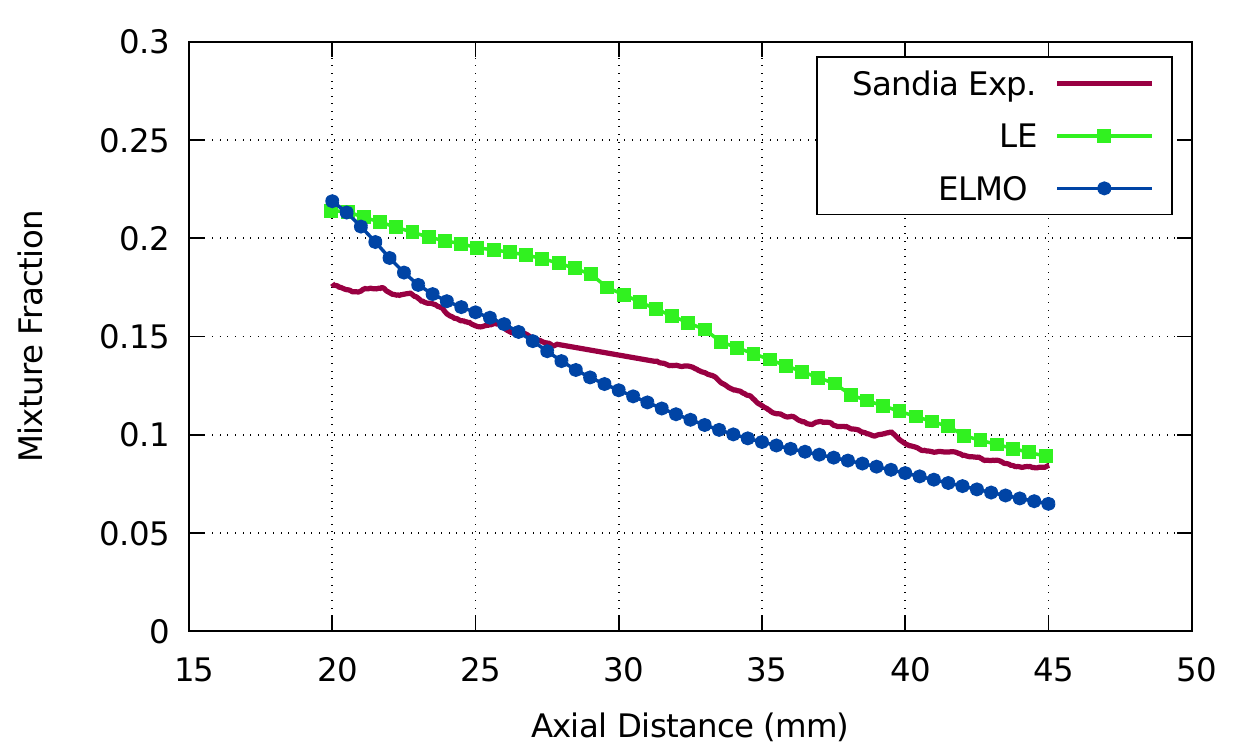} }}%
    \caption{Spray H simulation compared against experimental result of Sandia lab. a) vapor penetration, b) transverse, and c) longitudinal mixture fraction distribution.}%
    \label{fig:SpH}%
\end{figure}

The next validation case is the ECN Spray H \cite{idicheria2007}. The injected fuel is n-heptane, which is more volatile than dodecane in Spray A. Injection parameters can be found in the reference \cite{ECN}. The value of 24\textdegree is considered as the spray angle \cite{pickett2011}, while the nozzle coefficient of area is 0.86.  The initial turbulent kinetic energy and dissipation are 5.02E-4 $m^2/s^2$ and 0.0176 $m^2/s^3$ respectively. The ELMO results along with the Lagrangian Eulerian solution are shown in figure \ref{fig:SpH} including vapor penetration (a), transverse (b) and longitudinal fuel distribution (c) at 1.13 ms after the start of injection. ELMO outperformed LE in terms of vapor penetration prediction. About fuel mixture fraction, ELMO prediction is consistent with the Lagrangian Eulerian solution. Again, these experimental data are located far beyond the point where ELMO capsules are terminated, and so other factors may play a role. Because Spray H is more volatile and the gas temperature so high, the capsules vaporize quickly and the transition to parcels occurs only at very early times, near the start of injection.  For the large majority of the simulation, the fuel vaporizes directly and completely from the capsules. 

The final test for the ELMO model is the gasoline spray G condition \cite{ECN,payri2019}. This eight-hole gasoline direct injection uses iso-octane under non-flashing conditions.  This test case is fundamentally different than the two previous, not only because this is a multi-hole injector, but because of the far lower ambient density. These gasoline direct injection conditions do not correspond to the diesel-relevant conditions under which Siebers composed the mixing-limited hypothesis.  There is no existing theoretical boundaries that define when the mixing-limited hypothesis is applicable and when it is not, so this test case provides empirical illumination of the model applicability. 

\begin{figure}[ht]
    \centering
    a) \includegraphics[width=0.9\linewidth] {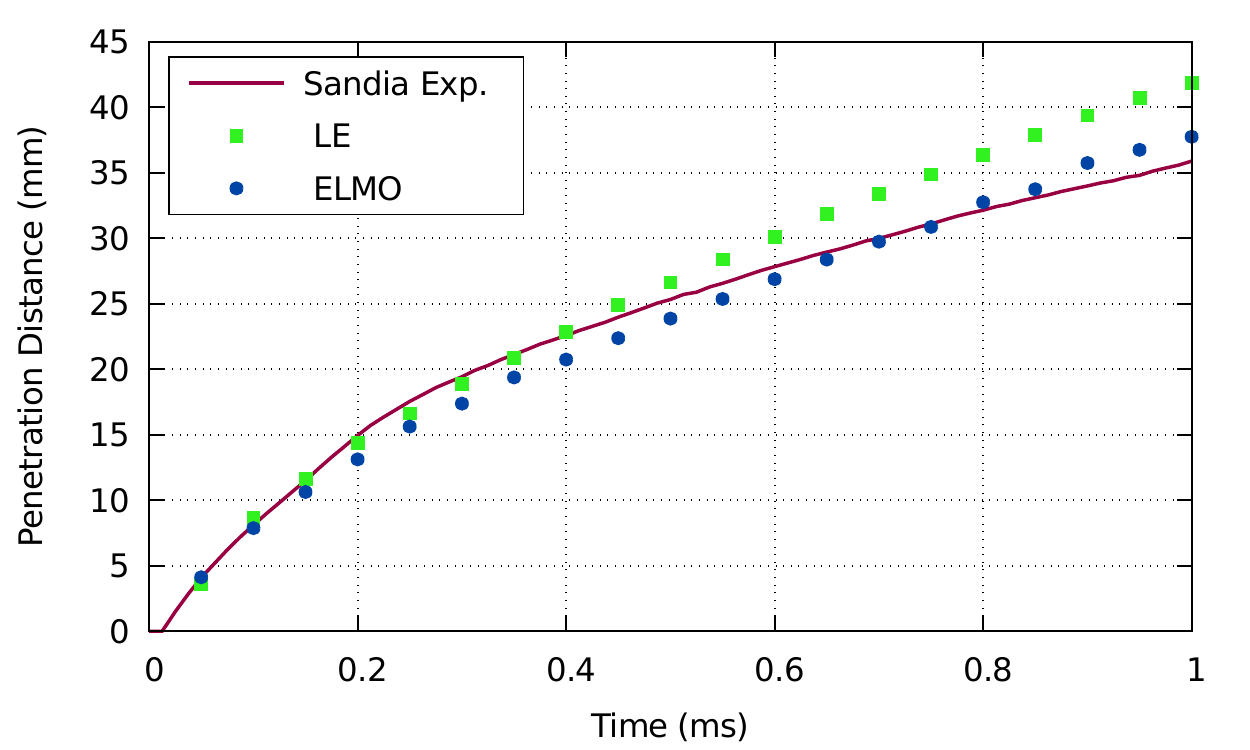}
    b) \includegraphics[width=0.9\linewidth]{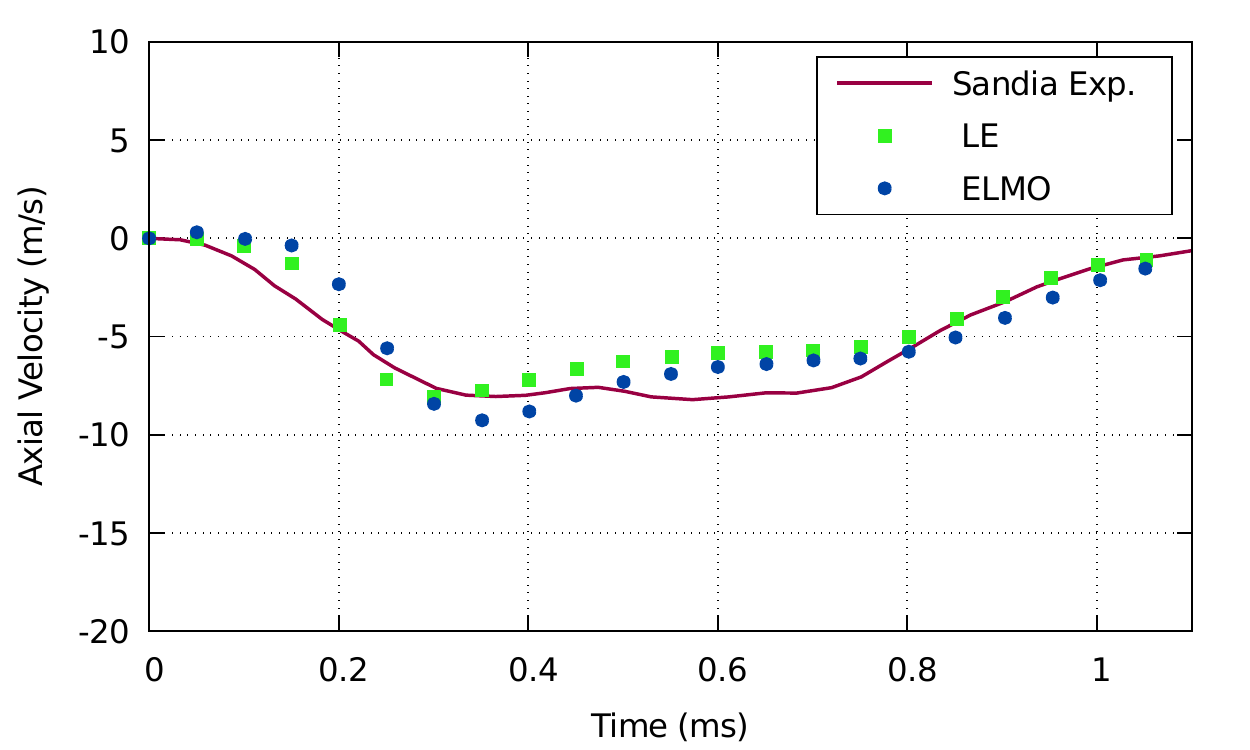}
    \caption{Spray G simulation compared against experimental result of Sandia lab. a) vapor penetration measured in the direction of the injector axis, b) Axial gas velocity at the centerline located at a downstream position of z = 15 $\mathrm{mm}$.  Positive values indicate motion away from the injector and negative values indicate motion towards the injector.}%
    \label{fig:SpG}%
\end{figure}

The two most critical spray input values are the spray cone angle = 20\textdegree and nozzle angle = 34\textdegree, taken from the work of Payri et al. \cite{payri2019}, and the coefficient of area, $C_a = 0.68$, from Moulai et al. \cite{moulai2015}. The initial turbulent kinetic energy and dissipation are 6.4E-3 $m^2/s^2$ and 0.486 $m^2/s^3$ respectively. The eight hole injection was reduced to a half sector with four nozzles. For validation, we use vapor penetration data from the Engine Combustion Network and measured gas velocity data from Sphicas et al. \cite{sphicas2017}.  The penetration data are measured in the direction of the injector axis, not the hole axis, as a function of time. The gas velocity data come from transient measurements made between plumes, at a point 15 $mm $ below the injector tip. Figure \ref{fig:SpG} provides experimental validation for the predicted penetration (a) and the axial gas velocity at a point on the axis of the injector (b).  The vapor penetration plot shows a good agreement between ELMO and experimental data. The axial velocity validation is based on the observation that the downward spray motion induces a temporary upward motion of air along the injector axis.  These velocity data show that shortly after the start of injection, the ambient gas begins moving towards the injector. In this test, the ELMO model predicts the right pattern matching the experimental data, similar to the LE simulation.

\section*{Conclusions}
 A mixing-oriented spray model has been conceived, implemented, and tested. ELMO is based on ideas of thermodynamic and inertial equilibrium in the dense spray core, resulting in an approach that targets mixing-limited conditions. ELMO, as demonstrated by the results generated here, can be applied to diesel sprays and gasoline sprays without adjustment. Compared to the drop-oriented models, ELMO has few arbitrary physical constants, providing little temptation to tune the model.  This makes the predictions reproducible and predictive. The computational cost of the model was roughly equivalent to traditional Lagrangian/Eulerian simulation.  In these non-combusting cases, the spray represents roughly eight percent or less of the total computational cost, so the cost of the ELMO model is not of great concern. 

 This work represents an attempt to make a fully-coupled, mixing-limited, CFD computation. The comparison with Lagrangian-Eulerian spray modeling is a particular challenge; given that Lagrangian-Eulerian models have been optimized and tuned in the last two decades. The fact that the ELMO model produces comparable accuracy in this initial effort is encouraging.

\section*{Acknowledgements}
 Funding for the project was provided by the Spray Combustion Consortium of automotive industry sponsors, including Convergent Science Inc.\, Cummins Inc.\, Ford Motor Co.\, Hino Motors Ltd.\, Isuzu Motors Ltd.\, Groupe Renault, and Toyota Motor Co. Experiments were performed at Combustion Research Facility, Sandia National Laboratory with support from the U.S. DOE Office of Vehicle Technologies. Sandia National Laboratories is a multi-mission laboratory managed and operated by National Technology and Engineering Solutions for Sandia LLC, a wholly owned subsidiary of Honeywell International, Inc.\, for the U.S. Department of Energy's National Nuclear Security Administration under contract DE-NA0003525. The authors also thank
Dr.\ Yajuvendra Shekhawat of Convergent Science Inc.\ for his help in using the CONVERGE code's full capabilities.


\bibliographystyle{ilass}
\bibliography{ilass} 


 \end{document}